\documentclass[onecolumn,showkeys]{revtex4}
\usepackage{graphicx}
\usepackage{dcolumn}
\usepackage{color}
\usepackage{bm}
\usepackage{amsmath}
\usepackage{amsfonts}
\usepackage{amssymb}
\RequirePackage[colorlinks,citecolor=blue,urlcolor=magenta,linkcolor=blue]{hyperref}
\input epsf
\begin{document}

\title{Spherically symmetric brane spacetime with bulk 
$f(\mathcal{R})$ gravity}

\author{Sumanta Chakraborty 
\footnote{sumantac.physics@gmail.com}
\footnote{sumanta@iucaa.ernet.in}}

\affiliation{IUCAA, Post Bag 4, Ganeshkhind,
Pune University Campus, Pune 411 007, India}

\author{Soumitra SenGupta 
\footnote{tpssg@iacs.res.in}}

\affiliation{Department of Theoretical Physics,
Indian Association for the Cultivation of Science,
Kolkata-700032, India}

\date{\today}
\begin{abstract}

Introducing $f(\mathcal{R})$ term in the five-dimensional bulk action 
we derive effective Einstein's 
equation on the brane using Gauss-Codazzi equation. This effective equation 
is then solved for different conditions on dark 
radiation and dark pressure to obtain various spherically symmetric solutions. 
Some of these static spherically 
symmetric solutions correspond to black hole 
solutions, with parameters induced from the bulk. Specially, the dark pressure and dark 
radiation terms (electric part of Weyl curvature) affect the 
brane spherically symmetric solutions significantly. We have solved for one parameter group 
of conformal motions where the dark radiation and dark pressure terms 
are exactly obtained exploiting the corresponding Lie symmetry. Various thermodynamic 
features of these spherically symmetric space-times are studied, 
showing existence of second order phase transition. 
This phenomenon has its origin in the higher curvature term with $f(\mathcal{R})$ gravity 
in the bulk. 

\end{abstract}


\maketitle
\section{Introduction}\label{Vacfintro}

Our four dimensional world might be embedded in a five dimensional space-time was proposed in 
\cite{Randall99} in order to explain the observed hierarchy between Electroweak and Planck scale.
Such extra dimensional models also have their origin in some suitable compactifications of  
ten dimensional $E_{8}\times E_{8}$ heterotic string theory \cite{Witten96}.

This scenario has attracted considerable attraction due to its elegant nature and simplicity. 
In this brane world scenario the standard model fields are confined on a 3-brane, while gravity 
can propagate both in the brane and the bulk. A single 3-brane, which is embedded 
in a five dimensional bulk has the five dimensional line element, 
$ds^{2}=e^{-A(y)}\eta _{\mu \nu}dx^{\mu}dx^{\nu}+dy^{2}$. The warp factor $e^{-A(y)}$ 
can be tuned properly to induce Einstein gravity on the brane as a leading order term. 
We could have also considered a two brane system, which comes with an additional 
field known as radion, representing separation between the branes, 
with interesting features \cite{Csaki2000,Chakraborty2014b}. 
However we will restrict ourselves only to the single 
brane system for the rest of the discussion.

However due to the presence of extra dimensions, we should expect deviation from 
Einstein theory, which play a significant role at high energies \cite{Maartens01,Maeda00}. 
Gravity sector also gets modified at electroweak scale $\sim 1$ TeV, changing the 
cosmological implications, which have been extensively studied in Ref. \cite{Harko01}. The 
effect of extra dimension on formation of black hole has been 
studied in Ref. \cite{Ghosh01}. Also these models have very interesting properties 
from the point of view of particle phenomenology 
\cite{Djouadi2008,Hundi2013,Chakraborty2014c,Dudas2005,Lukas1999}.

In General Relativity the exterior space-time of a spherically 
symmetric black hole or a compact object is standard 
Schwarzschild geometry. However due to the presence of an extra dimension in the brane world 
scenario the Schwarzschild solution gets modified non-trivially. This originates due to 
high energy corrections, Weyl stress on gravitons propagating in the bulk. One such solution
was obtained in Ref. \cite{Dadhich00}, in the form of Reissner-N\"{o}rdstrom solution. 
The interior solution can be matched to a brane world star having constant 
energy density \cite{Germani01,Casadio14,Ovalle13}. A non singular solution for black holes in these models 
can be obtained by relaxing the condition of zero scalar curvature 
while retaining null energy condition \cite{Dadhich03,Casadio12}. Also the Gauss-Codazzi equations can be 
solved in Randall-Sundrum type II model to get exterior solution for spherically symmetric star 
\cite{Visser03}. The various classes of vacuum solutions has been obtained in Ref. \cite{Harko04} 
by solving the vacuum field equations on the brane obtained from Gauss Codazzi equation. The 
results of various such calculations suggest that brane world black hole horizons has the 
peculiar structure of a "pancake". 

In recent years, there has been a new concept in General Relativity suggesting modifications 
of Einstein-Hilbert action in order to explain the late time cosmic acceleration to inflation. 
This is achieved by introducing higher curvature terms in the action, and a very promising 
candidate among such modifications is $f(\mathcal{R})$ gravity theories 
(for recent reviews see \cite{Nojiri11}). 
The main difficulties with these modifications are, they become infected with ghost modes. 
However $f(\mathcal{R})$ theory on a constant curvature hyper surface is shown 
to be ghost free \cite{Nojiri07}. 
The modification due to introduction of $f(\mathcal{R})$ term in the Lagrangian can address 
variety of problems e.g. four cosmological phases \cite{Nojiri06}, 
late time cosmic acceleration \cite{Bamba08}, 
initial power law inflation \cite{Star80}, rotation curves of spiral galaxies \cite{Troisi06}, 
detection of 
gravitational waves \cite{Corda08} and many others. This theory has also the potential 
to pass through all known tests of general relativity.

Motivated by such striking properties of $f(\mathcal{R})$ gravity it is also introduced in 
brane world models, where the five dimensional action is modified by introduction 
of $f(\mathcal{R})$ term in the bulk, with $\mathcal{R}$ being the Ricci scalar of the 
five dimensional theory. In particular for bulk geometry with high 
curvature $\sim$ Planck scale, such higher order corrections 
to gravity are expected to become extremely relevant. Effective gravitational equations on the brane
have been obtained in Ref. \cite{Silva13,Bazeia13,Haghani2012,Borzou2009} while perturbations 
on the scalar and tensor 
modes on the brane has been studied in Ref. \cite{Parry05}. Cosmology on these brane world 
models Ref. \cite{Bal10} along with brane world sum rules have also been discussed in these 
$f(\mathcal{R})$ gravity models \cite{Silva11}. The nature of warped geometric 
models in this $f(\mathcal{R})$ gravity theory with constant bulk curvature has been 
obtained in Ref. \cite{Chakraborty14} and the graviton KK mode masses in these models have 
been examined in the light of recent ATLAS data in LHC. 

Ever since the pioneering works of Regge and Wheeler \cite{Regge57,Zerilli70,Vish70}, the stability of a four dimensional black hole under linear perturbation 
has been investigated extensively. The importance of linear stability of a black hole 
can be understood as follows: the black hole solutions should describe 
the final state of gravitational collapse and thus they should be stable against small fluctuations. Also 
technically, this implies that at the order of linear perturbation, Einstein equation reduces to 
a simple set of wave equations. For the static situation, these equations resemble Schr\"{o}dinger 
equation with time dependent Hamiltonian. Thus the stability analysis becomes equivalent to 
a simple, quantum mechanical problem. We also mention that there are solutions 
which describe naked singularity, and stability of a naked singularity is an important issue 
from the viewpoint of cosmic censorship conjecture. In this work we will use the wave equations 
to study the stability \cite{Kodama2003a,Kodama2003b,Kodama2004,Giveon2004}.

An important aspect of black hole physics, pioneered 
by Bekenstein, shows a remarkable similarity between black hole and a thermodynamic system. 
The similarity arises from the fact that just like a thermodynamic 
system one can attribute temperature to a  black hole 
(known as Hawking temperature) which is 
proportional to the surface gravity and also an entropy proportional to the horizon area 
\cite{Bekenstein73,Hawking73,Hawking75,Padmanabhan2005a,Wald2001,Padmanabhan2010a}. 
Any arbitrary black 
hole can be characterized by three parameters, its mass, charge and angular momentum. 
The thermodynamic stability of such a system can be determined by the sign of 
heat capacity just like any normal thermodynamic system. For a black hole 
the criteria $c_{v}<0$ makes the system thermodynamically unstable. However if the 
specific heat changes sign as well as diverges in its parameter space, then 
it indicates a second order phase transition \cite{Davies77,Davies89}. 
Phase transitions in various black hole solutions have been studied extensively 
in Einstein gravity as well as in alternative gravity 
theories \cite{Page83,Hut77,York86,Myung07,Chakraborty10,Sengupta13}. 

The purpose of this work is to consider various spherically symmetric vacuum space-times on 
the brane obtained from $f(\mathcal{R})$ action on the bulk. 
In order to achieve this we consider the decomposition of electric part of the  
Weyl tensor into dark radiation and dark pressure terms. It turns out that these  
determine the space-time geometry we are considering. Moreover some simple integrability conditions 
leads to different classes of vacuum solutions. 
These issues are addressed in Sec. \ref{Vacfstat} and Sec. \ref{Vacfclass}. Then 
we have discussed stability of black holes and naked singularities in these spacetime 
in Sec. \ref{Vacfstab}.

Next we consider vacuum space-time related to Lie groups of transformation. As a 
simple situation we consider spherically symmetric and static solutions with 
the metric tensor admitting one parameter group of conformal motion. With proper 
integrability condition an exact solution corresponding to a brane with one parameter 
group of motions can be obtained (see Sec. \ref{statcon}). 

Finally we consider the thermodynamics of these black hole solutions. 
As these solutions are  induced on the brane due to bulk action, the thermodynamic properties 
are related to the dark pressure and radiation terms coming  
from the electric part of Weyl tensor and thus the 
thermodynamic properties of the brane black holes are directly related to those 
of bulk space-time (see Sec. \ref{Vacftherm}). We finally conclude with a discussion 
on our results. 

\section{Static, Spherically Symmetric Field Equations On The Brane}\label{Vacfstat}

To obtain the vacuum solution we start from the bulk action with 
$f(\mathcal{R})$ term as,
\begin{equation}\label{eqsfo1}
S=\int d^{5}x\sqrt{-g}\left[f(\mathcal{R})+\mathcal{L}_{m}\right]
\end{equation}
where $\mathcal{L}_{m}$ is the matter Lagrangian, 
$g_{AB}$ is the bulk metric and $\mathcal{R}$ is the bulk Ricci scalar. 
The bulk indices $A,B$ runs through $0\ldots 4$ i.e. over all the 
space-time dimensions.
The variation of the action $S$ with respect to bulk metric $g_{AB}$ leads to,
\begin{equation}\label{eqsfo2}
f'(\mathcal{R})\mathcal{R}_{AB}-\frac{1}{2}g_{AB}f(\mathcal{R})+g_{AB}\square f'(\mathcal{R})-\nabla _{A}\nabla _{B}f'(\mathcal{R})
=\kappa _{5}^{2}T_{AB}
\end{equation}
Here the negative vacuum energy density $\Lambda$ on the bulk 
and the brane energy-momentum tensor are the sources of the gravitational field.
Eq. (\ref{eqsfo2}) can be put into the form,
\begin{eqnarray}\label{eqsfo3}
G_{AB}&\equiv & \mathcal{R}_{AB}-\frac{1}{2}\mathcal{R}g_{AB}=T_{AB}^{tot}
\nonumber
\\
T_{AB}^{tot}&=&\frac{1}{f'(\mathcal{R})}\left[\kappa _{5}^{2}T_{AB}
-\left(\frac{1}{2}\mathcal{R}f'(\mathcal{R})-
\frac{1}{2}f(\mathcal{R})+\square f'(\mathcal{R})\right)g_{AB}+
\nabla _{A}\nabla _{B}f'(\mathcal{R})\right]
\nonumber
\\
T_{AB}&=&-\Lambda g_{AB}+
\delta (y)\left(-\lambda _{T} h_{\mu \nu}+\tau _{\mu \nu}\right)\delta ^{\mu}_{A}\delta ^{\nu}_{B}
\end{eqnarray}
where $\tau _{\mu \nu}$ is the brane energy-momentum tensor 
and $\lambda _{T}$ is the corresponding brane tension. 
Also the quantity $h_{\mu \nu}$ is the induced metric on 
$y=\textrm{constant}$ hypersurfaces.\\
The effective four-dimensional 
gravitational equations on the brane are,
\begin{equation}\label{eqsfo4}
G_{\mu \nu}=-\Lambda _{4}h_{\mu \nu}+
8\pi G_{N}\tau _{\mu \nu}+\kappa _{5}^{2}\pi _{\mu \nu}+Q_{\mu \nu}-
E_{\mu \nu}
\end{equation}
where, 
\begin{eqnarray}\label{eqsfo5}
\Lambda _{4}&=&\frac{1}{2}\kappa _{5}^{2}\left(\frac{\Lambda}{f'(\mathcal{R})}+
\frac{1}{6}\kappa _{5}^{2}\lambda ^{2}\right)
\\
G_{N}&=&\frac{\kappa _{5}^{4}\lambda}{48\pi}
\\
\pi _{\mu \nu}&=&-\frac{1}{4}\tau _{\mu \alpha}\tau _{\nu}^{\alpha}+
\frac{1}{12}\tau \tau _{\mu \nu}+
\frac{1}{8}h_{\mu \nu}\tau _{\alpha \beta}\tau ^{\alpha \beta}-
\frac{1}{24}h_{\mu \nu}\tau ^{2}
\\
Q_{\mu \nu}&=&\left[g(\mathcal{R})h_{\mu \nu}+
\frac{2}{3}\frac{\nabla _{A}\nabla _{B}f'(\mathcal{R})}{f'(\mathcal{R})}
\left(h_{\mu}^{A}h_{\nu}^{B}+
n^{A}n^{B}h_{\mu \nu}\right) \right]_{y=0}
\label{Qexp}
\end{eqnarray}
with,
\begin{equation}\label{eqsfo6}
g(\mathcal{R})\equiv \frac{1}{4}\frac{f(\mathcal{R})}{f'(\mathcal{R})}-
\frac{1}{4}\mathcal{R}-\frac{2}{3}\frac{\square f'(\mathcal{R})}{f'(\mathcal{R})}
\end{equation}
Note that for $f(\mathcal{R})=\mathcal{R}$, we retrieve the usual Gauss-Codazzi equation 
for a pure Einstein gravity in the bulk. We now proceed to simplify the expression for $Q_{\mu \nu}$. 
The normal to $y=\textrm{constant}$ hypersurface being $n_{A}=\partial _{A}y$, 
we have $n_{\mu}=0$. In addition if we assume that $\partial _{\mu}\mathcal{R}=0$ then using 
the relations: $\nabla _{A}\nabla _{B}f'(\mathcal{R})=$ 
$f''(\mathcal{R})\nabla _{A}\nabla _{B}\mathcal{R}
+f'''(\mathcal{R})\nabla _{A}\mathcal{R}\nabla _{B}\mathcal{R}$ 
and $\nabla _{A}\mathcal{R}\nabla _{B}\mathcal{R}h^{A}_{\mu}h^{B}_{\nu}=$ 
$\nabla _{\mu}\mathcal{R}\nabla _{\nu}\mathcal{R}
-\nabla _{\mu}\mathcal{R}\nabla _{B}\mathcal{R}n^{B}n_{\nu}$ 
$-\nabla _{A}\mathcal{R}\nabla _{\nu}\mathcal{R}n^{A}n_{\mu}$ 
$-\nabla _{A}\mathcal{R}\nabla _{B}\mathcal{R}n^{A}n^{B}n_{\mu}n_{\nu}$ 
along with a similar expression for 
$\nabla _{A}\nabla _{B}\mathcal{R}h^{A}_{\mu}h^{B}_{\nu}$ Eq. (\ref{Qexp}) reduces to,
\begin{equation}
Q_{\mu \nu}=\left(g(\mathcal{R})
+\frac{2}{3}\frac{\nabla _{A}\nabla _{B}f'(\mathcal{R})}
{f'(\mathcal{R})}n^{A}n^{B}\right)_{y=0}h_{\mu \nu}
\equiv F(\mathcal{R})h_{\mu \nu}
\end{equation}
Now the scalar curvature for the bulk must be a well behaved quantity, 
and we can expand it in a Taylor series around $y=0$ 
hypersurface, as, 
$\mathcal{R}=\mathcal{R}_{0}+\mathcal{R}_{1}y+\mathcal{R}_{2}y^{2}/2+\mathcal{O}(y^{3})$. Since 
bulk curvature depends only on the extra dimension $y$, all the coefficients are constants. 
Thus all the derivatives 
calculated at $y=0$ yield a constant contribution which 
does not depend on any of the brane coordinates. 

The electric part of the Weyl tensor $E_{\mu \nu}$ has its origin in the
nonlocal effect from free bulk gravitational field. This is 
the projection of bulk Weyl tensor such that, $E_{AB}=C_{ABCD}n^{C}n^{D}$ along with 
$E_{AB}=E_{\mu \nu}\delta ^{\mu}_{A}\delta ^{\nu}_{B}$ on the brane ($y \rightarrow 0$). 
From the Gauss-Codazzi equation we also have conservation 
of energy momentum tensor as, $D_{\mu}T^{\mu \nu}=0$, 
where $D_{\mu}$ is the brane covariant derivative. This also imposes 
restrictions on projected Weyl tensor from Bianchi identities. 
Following Ref. \cite{Maartens01} the projected Weyl tensor 
can be expanded as,
\begin{equation}\label{eqsfo8}
E_{\mu \nu}=-k^{4}\left[U(r)\left(u_{\mu}u_{\nu}
+\frac{1}{3}\xi _{\mu \nu}\right)+P_{\mu \nu}+2Q_{(\mu}u_{\nu )} \right]
\end{equation}
with $k=k_{5}/\sqrt{8\pi G_{N}}$ and $\xi _{\mu \nu}=h_{\mu \nu}+u_{\mu}u_{\nu}$. 
This decomposition is with respect to the four velocity field $u_{\mu}$. 
The respective terms in the above expression are, 
the "Dark Radiation" term, $U=-\frac{1}{k^{4}}E_{\mu \nu} u^{\mu}u^{\nu}$, which is a 
scalar, 
$Q_{\mu}=\frac{1}{k^{4}}\xi _{\mu}^{\alpha}E_{\alpha \beta}$ is a spatial vector 
and $P_{\mu \nu}=-\frac{1}{k^{4}}\left[\xi ^{\alpha}_{(\mu}\xi ^{\beta}_{\nu )}-
\frac{1}{3}h_{\mu \nu}h^{\alpha \beta} \right]E_{\alpha \beta}$ 
is a spatial, trace free, symmetric tensor. 
For static solutions, $Q_{\mu}=0$, while the constraint becomes
dependent on dark radiation $U(r)$, vector $A_{\mu}=A(r)r_{\mu}$ and a tensor 
$P_{\mu \nu}=P(r)\left(r_{\mu}r_{\nu}-\frac{1}{3}\xi _{\mu \nu} \right)$. Here $r_{\mu}$ 
is unit radial vector.

In order to obtain solution in a source free region on the brane, 
brane energy momentum tensor 
appearing on the right hand side of effective Einstein's equation 
is taken to be zero. Thus we readily obtain $\tau _{\mu \nu}=0=\pi _{\mu \nu}$. 
Also from the previous discussion it is evident that
$\mathcal{R}$ is dependent only on $y$ and 
on the brane (at $y=0$) all its derivatives with respect to coordinates 
become constants. Then the Einstein equation becomes,
\begin{equation}\label{eqsfo7}
G_{\mu \nu}=-\Lambda _{4}h_{\mu \nu}+F(\mathcal{R})h_{\mu \nu}-E_{\mu \nu}
\end{equation}
Now we choose an ansatz for spherically symmetric solution 
in the form,
\begin{equation}\label{eqsfo9}
ds^{2}=-e^{\nu (r)}dt^{2}+e^{\lambda (r)}dr^{2}+r^{2}d\Omega ^{2}
\end{equation}
For this choice the effective Einstein's equation and energy-momentum conservation equation 
on the brane become,
\begin{eqnarray}
-e^{-\lambda}\left(\frac{1}{r^{2}}-\frac{\lambda '}{r} \right)+
\frac{1}{r^{2}}&=&\left(\Lambda _{4}-F(\mathcal{R})\right)+\frac{3}{4\pi G\lambda _{T}}U
\label{eqsfo10}
\\
e^{-\lambda}\left(\frac{\nu '}{r}+\frac{1}{r^{2}}\right)-\frac{1}{r^{2}}&=&F(\mathcal{R})-\Lambda _{4}
+\frac{1}{4\pi G\lambda _{T}}(U+2P)
\label{eqsfo11}
\\
e^{-\lambda}\left(\nu ''+\frac{\nu '^{2}}{2}+\frac{\nu '-\lambda '}{r}
-\frac{\nu '\lambda '}{2} \right)&=&
2\left(F(\mathcal{R})-\Lambda _{4}\right)+\frac{1}{2\pi G\lambda _{T}}(U-P)
\label{eqsfo12}
\\
\nu '&=&-\frac{U'+2P'}{2U+P}-\frac{6P}{r(2U+P)}
\label{eqsfo13}
\end{eqnarray}
where we have denoted $a'\equiv da/dr$. 
Now Eq. (\ref{eqsfo10}) can be solved for $e^{-\lambda}$ to yield,
\begin{equation}\label{eqsfo14}
e^{-\lambda}=1-\frac{\Lambda _{4}-F(\mathcal{R})}{3}r^{2}-\frac{Q(r)}{r}-\frac{C_{1}}{r}
\end{equation}
where $C_{1}$ is an arbitrary constant of integration. 
The quantity $Q(r)$ is defined as,
\begin{equation}\label{eqsfo15}
Q(r)=\frac{48\pi G}{k_{4}^{4}\lambda _{b}}\int r^{2}U(r)dr
\end{equation}
We can interpret the term $Q$ as equivalent to gravitational mass originating from 
dark radiation and henceforth will be referred as dark mass. 
In the limit $f(\mathcal{R})\rightarrow \mathcal{R}$, $\Lambda _{4}\rightarrow  0$ as well 
as $U\rightarrow 0$ we retrieve the standard Schwarzschild 
solution. This helps us to identify the arbitrary constant as $C_{1}=2GM$, $M$ 
being the constant mass of the gravitating body. Also we can obtain the differential 
equations that are satisfied by dark radiation $U(r)$ and dark pressure $P(r)$ 
in static spherically symmetric space-time. Eliminating 
$\nu '$ from Eq. (\ref{eqsfo13}) and Eq. (\ref{eqsfo11}) and using 
$e^{-\lambda}$ from Eq. (\ref{eqsfo14}) 
we obtain:
\begin{eqnarray}
\frac{dU}{dr}&=&-2\frac{dP}{dr}-6\frac{P}{r}
-\frac{(2U+P)\left[2GM+Q+\left\lbrace \alpha (U+2P)+2\chi /3\right\rbrace r^{3}\right]}
{r^{2}\left(1-\frac{2GM}{r}-\frac{Q(r)}{r}-\frac{\Lambda _{4}-F(\mathcal{R})}{3} r^{2} \right)}
\label{eqsfo16}
\\
\frac{dQ}{dr}&=&3\alpha r^{2}U
\label{eqsfo17}
\end{eqnarray}
where we introduce two extra parameters, 
$\alpha =\left(1/4\pi G\lambda _{T}\right)$ 
and $\chi =F(\mathcal{R})-\Lambda _{4}$.
Now we define the following quantities in order 
to transform the above differential equation into a more 
convenient form which will be used extensively later,
\begin{equation}\label{eqsfo18}
q=\frac{2GM+Q}{r};~\mu =3\alpha r^{2}U;~p=3\alpha r^{2}P;~\theta =\ln r;~2\chi r^{2}=\ell
\end{equation}
In terms of these variables  
the differential equations satisfied by the dark 
radiation and dark pressure are,
\begin{eqnarray}
\frac{dq}{d\theta}&=&\mu -q
\label{eqsfo19}
\\
\frac{d\mu}{d\theta}&=&-\left(2\mu +p \right)\frac{q+\frac{1}{3}\left( \mu +2p\right)+\frac{\ell}{3}}{1-q+\frac{\ell}{6}}-2\frac{dp}{d\theta}+2\mu -2p
\label{eqsfo20}
\end{eqnarray}
Thus the Eqs. (\ref{eqsfo10}) to (\ref{eqsfo13}) are the effective field equations, on 
the brane, while the Eqs. (\ref{eqsfo19}) to (\ref{eqsfo20}) represent 
equations for the source terms in the bulk i.e. dark pressure and dark radiation.

\section{Various Classes of Solutions On The Brane}\label{Vacfclass}

Eqs. (\ref{eqsfo16}) and (\ref{eqsfo17}) can not be solved for dark radiation $U$ 
and dark pressure $P$ simultaneously unless we have a relation connecting them. 
We therefore choose some possible relations between the dark 
radiation $U$ and dark pressure $P$ which essentially define 
different equations of state. For different such choices we 
get different solutions. In this section we impose 
certain conditions on dark radiation $U$ and dark pressure $P$, 
to obtain the corresponding solution. It turns out that the solutions 
are very distinct for different choices.

\subsection{\textbf{Case-I.}$\mathbf{~U=0}$}

This condition comes with vanishing dark radiation, which imply readily $Q=0$. 
In this scenario, one of the metric elements can be given by,
\begin{equation}\label{clb09}
e^{-\lambda}=1+\frac{F(\mathcal{R})-\Lambda _{4}}{3}r^{2}-\frac{2GM}{r}
\end{equation}
The differential equation satisfied by the dark pressure $P(r)$ is given by,
\begin{equation}\label{clb10}
\frac{dP}{dr}+3\frac{P}{r}+\frac{P\left(GM+\alpha r^{3}P
+\left(F(\mathcal{R})-\Lambda _{4} \right)/3 r^{3} \right)}{r^{2}\left(1-\frac{2GM}{r}+\frac{F(\mathcal{R})-\Lambda _{4}}{3}r^{2}\right)}=0
\end{equation}
while the differential equation satisfied by $\nu$ is given by,
\begin{equation}\label{clb11}
\nu '=\frac{2\left(GM+\alpha r^{3}P
+\left(F(\mathcal{R})-\Lambda _{4} \right)/3 r^{3} \right)}{r^{2}\left(1-\frac{2GM}{r}+\frac{F(\mathcal{R})-\Lambda _{4}}{3}r^{2}\right)}
\end{equation}
Solution for these two differential equations give the pressure 
and metric for this case. Note that in this situation the metric element $e^{\nu}$ is solely 
determined from the pressure, which can be seen directly from Eq. (\ref{clb11}) and Eq. (\ref{clb10}) as, 
$\nu '=-2P'/P-6/r$. This equation can be integrated to yield, 
$\exp(\nu)=C_{2}/r^{6}P^{2}$, where $C_{2}$ is 
an arbitrary constant of integration. Thus once pressure equation is solved, 
the metric element is also known. 

In order to obtain the pressure two quantities $r_{1}$ and 
$d$ would be important with the following expressions:
\begin{eqnarray}\label{newclb02}
r_{1}&=&\frac{3^{-2/3}\left(F(\mathcal{R})-\Lambda _{4}\right)
+\left(-GM\left(F(\mathcal{R})-\Lambda _{4}\right)^{2}
+\sqrt{3}\sqrt{\frac{\left(F(\mathcal{R})-\Lambda _{4}\right)^{3}}{27}
+\left[-1+9\left(F(\mathcal{R})-\Lambda _{4}\right)G^{2}M^{2}\right]}\right)^{2/3}}
{3^{-1/3}\left(F(\mathcal{R})-\Lambda _{4}\right)\left(-GM\left(F(\mathcal{R})-\Lambda _{4}\right)^{2}
+\sqrt{3}\sqrt{\frac{\left(F(\mathcal{R})-\Lambda _{4}\right)^{3}}{27}
+\left[-1+9\left(F(\mathcal{R})-\Lambda _{4}\right)G^{2}M^{2}\right]}\right)^{1/3}}
\\
d&=&\frac{1}{\left(F(\mathcal{R})
-\Lambda _{4}\right)^{2}}\Big[-3^{5/6}\sqrt{\frac{\left(F(\mathcal{R})-\Lambda _{4}\right)^{3}}{27}
\left[-1+9\left(F(\mathcal{R})-\Lambda _{4}\right)G^{2}M^{2}\right]}
\nonumber
\\
&\times& \left(-GM\left(F(\mathcal{R})-\Lambda _{4}\right)^{2}
+\sqrt{3}\sqrt{\frac{\left(F(\mathcal{R})-\Lambda _{4}\right)^{3}}{27}
+\left[-1+9\left(F(\mathcal{R})-\Lambda _{4}\right)G^{2}M^{2}\right]}\right)^{1/3}
\nonumber
\\
&+&\frac{\left(F(\mathcal{R})-\Lambda _{4}\right)}{3}
\left(-3GM\left(F(\mathcal{R})-\Lambda _{4}\right)^{2}
+\sqrt{\left(F(\mathcal{R})-\Lambda _{4}\right)^{3}
\left[-1+9G^{2}M^{2}\left(F(\mathcal{R})-\Lambda _{4}\right)\right]} \right)^{2/3}
\nonumber
\\
&+&\frac{\left(F(\mathcal{R})-\Lambda _{4}\right)^{2}}{3}\left(1+3GM\left\lbrace-3GM\left(F(\mathcal{R})-\Lambda _{4}\right)^{2}
+\sqrt{\left(F(\mathcal{R})-\Lambda _{4}\right)^{3}
\left[-1+9G^{2}M^{2}\left(F(\mathcal{R})-\Lambda _{4}\right)\right]} 
\right\rbrace \right)^{1/3}\Big]
\nonumber
\end{eqnarray}
With these variables the solution for the pressure is obtained as:
\begin{eqnarray}\label{newclb01}
P(r)&=&h(r)\left[\int \frac{\alpha r^{3}}{r^{2}\left(1-2GM/r+(F(\mathcal{R}-\Lambda _{4})r^{2}/3)\right)}h(r)+C_{1}\right]^{-1}
\\
h(r)&=&\frac{1}{r^{3}}\left(\frac{1}{r}-\frac{1}{r_{1}}\right)^{-\frac{3GMA}{\left(F(\mathcal{R})-\Lambda _{4}\right)r_{1}d}}
\exp\left[-\frac{3GM\left(d-r_{1}^{2}\right)}{\left(F(\mathcal{R})-\Lambda _{4}\right)r_{1}^{2}d\left(1+d/2r_{1}^{2}\right)\sqrt{4d-r_{1}^{2}}}
\arctan \left(\frac{r_{1}+2d/r_{1}}{\sqrt{4d-r_{1}^{2}}}\right)\right]
\nonumber
\\
&\times&\left(1+\frac{r_{1}}{r}
+\frac{d}{r^{2}} \right)^{-\frac{3GM}{2\left(F(\mathcal{R})-\Lambda _{4}\right)r_{1}d\left(1+d/2r_{1}^{2}\right)}}
\end{eqnarray}
From the above expression it is evident that at $r=r_{1}$ 
the metric element $e^{\nu}$ vanishes. Thus the space-time has 
an event horizon located at $r=r_{1}$ 
with its characteristic thermodynamic features. 

\subsection{\textbf{Case-II.}$\mathbf{~P=0}$}

In this situation Eqs. (\ref{eqsfo19}) and (\ref{eqsfo20}) reduces to the following form,
\begin{eqnarray}
\frac{dq}{d\theta}=\mu -q
\label{clb12}
\\
\frac{d\mu}{d\theta}=2\mu \left[\frac{6-\ell -2\mu -12q}{6+\ell -6q} \right]
\label{clb13}
\end{eqnarray}
These two equations can be combined to yield a single differential equation such that,
\begin{equation}\label{clb14}
\left(6+\ell -6q\right)\frac{d^{2}q}{d\theta ^{2}}+\left(26q-6-\ell \right)\frac{dq}{d\theta}
+4\left(\frac{dq}{d\theta} \right)^{2}+2q\left(14q-6-\ell \right)=0
\end{equation}
The transformations $dq/d\theta =1/v$ and $v=w\left(6-6q+\ell \right)^{-2/3}$ 
lead to the following differential equation,
\begin{equation}\label{clb15}
\frac{dw}{dq}-\left(26q-6-\ell \right)\left(6-6q+\ell \right)^{-5/3}w^{2}-2q\left(14q-6-\ell \right)\left(6-6q+\ell \right)^{-7/3}w^{3}=0
\end{equation}
The above differential equation has a particular solution, 
$w=-\frac{1}{q}\left(6-6q+\ell \right)^{2/3}$. However for 
a wider class of solutions we define a new variable 
$\eta =\left(6-6q+\ell \right)^{-1/3}$. This leads to the differential equation,
\begin{equation}\label{clb16}
\frac{dw}{d\eta}-\frac{10\eta ^{3}+10/13 \ell \eta ^{3}-13/16}{\eta ^{2}}
+\frac{\left[\eta ^{3}\left(1+\ell /6\right)-1 \right]\left[7/3 -\eta ^{3}\left(10+4\ell /3 \right) \right]}{\eta ^{3}}w^{3}=0
\end{equation}
It is hard to find an exact solution of this differential equation. Therefore 
we resort to approximated methods. For that purpose we choose the differential 
equation (\ref{clb14}) and making Laplace transform of this equation we get,
\begin{eqnarray}
\mathcal{L}\left(\left[3+\chi e^{2\theta}\right]\frac{d^{2}q}{d\theta ^{2}}
-\left[3+\chi e^{2\theta} \right]\frac{dq}{d\theta}-4q\left[3+\chi e^{2\theta} \right] \right)
\nonumber
\\
=\mathcal{L}\left(3q\frac{d^{2}q}{d\theta ^{2}}-13q\frac{dq}{d\theta}
+4\left(\frac{dq}{d\theta} \right)^{2}-14q^{2}\right)
\end{eqnarray} 
Then using the convolution theorem in the form,
\begin{equation}
\mathcal{L}^{-1}\left(\tilde{f}(s)\tilde{g}(s)\right)=\int _{a}^{b}f(t-u)g(u)du
\end{equation}
we readily obtain the following integral solution,
\begin{equation}
q(\theta)=q_{0}(\theta)+\int _{\theta _{0}}^{\theta}f(\theta -y)\left[3q\frac{d^{2}q}{d\theta ^{2}}-13q\frac{dq}{d\theta}
+4\left(\frac{dq}{d\theta} \right)^{2}-14q^{2} \right]dx
\end{equation}
where we have the following functions,
\begin{eqnarray}
f(x-y)=\frac{1}{9}\left(e^{2(x-y)}-e^{-(x-y)} \right)
\\
q_{0}(\theta)=A_{1}e^{-\theta}+A_{2}e^{2\theta}
\\
A_{1}=\left[\left(3q_{0}-\mu _{0}\right)+\left((3+2\chi)q_{0}-\mu _{0}\right)\right]e^{\theta _{0}}/3
\\
A_{2}=\mu (\theta _{0})e^{-2\theta _{0}}/3
\end{eqnarray}
Having obtained an integral solution we now move forward to determine the metric. 
However the solution is usually obtained by successive approximation methods, which invokes 
iterations. At zeroth order we get the solution by using only the linear part of the differential 
equation (\ref{clb14}) and will be denoted by $q_{0}$. Then we can write our full solution as 
a limiting process, such that $q(\theta)=\lim _{m \rightarrow \infty} q_{m}(\theta)$. In this situation 
for $m\in N$, we have the iterative solution at m-th order 
connected to $(m-1)$ th order by the following integral equation,
\begin{equation}
q_{m}(\theta)=\int _{\theta _{0}}^{\theta} F(\theta -y)\left[3q_{m-1}\frac{d^{2}q_{m-1}}{d\theta ^{2}}-13q_{m-1}\frac{dq_{m-1}}{d\theta}
+4\left(\frac{dq_{m-1}}{d\theta} \right)^{2}-14q_{m-1}^{2} \right]dy+q_{m-1}(\theta)
\end{equation}
Then following Ref. \cite{Harko04} the zeroth order static and spherically symmetric solution 
to the field equations turn out to be,
\begin{eqnarray}
e^{\nu}=C_{0}\sqrt{\frac{\alpha}{A_{2}}}
\label{new01}
\\
e^{-\lambda}=1-\frac{A_{1}}{r}-A_{2}r^{2}
\label{new02}
\\
U=\frac{A_{2}}{\alpha}
\label{new02a}
\end{eqnarray}
where we have $C_{0}$ as an arbitrary integration constant. After 
using one more iteration i.e. upto first order approximation the metric components 
are obtained as,
\begin{eqnarray}
e^{\nu}&=&C_{0}\sqrt{\frac{\alpha r_{0}}{2}}\sqrt{\frac{r}{A_{2}(r_{0}-r)[A_{1}+A_{2}rr_{0}^{2}+A_{2}r_{0}r^{2}]}}
\label{new03}
\\
e^{-\lambda}&=&1+\frac{A_{2}r_{0}^{2}[(4A_{2}r_{0}^{2}/5)+
A_{1}]}{r}-3A_{1}A_{2}r-2A_{2}\left(2A_{2}r_{0}^{2}-A_{1}/r_{0} \right)r^{2}
+6A_{2}^{2}r^{4}/5
\label{new04}
\end{eqnarray}
Note that the dependence on $f(\mathcal{R})$ gravity appears through the $A_{1}$ factor. However 
the dependence is quiet complicated and affects both the metric elements.

\subsection{\textbf{Case-III.}$\mathbf{~2U+P=0}$}

For this choice Eq. (\ref{eqsfo16}) yields,
\begin{eqnarray}
P(r)&=&\frac{P_{0}}{r^{4}}
\label{clb01}
\\
U(r)&=&-\frac{P_{0}}{2r^{4}}
\label{clb02}
\end{eqnarray}
where $P_{0}$ is an arbitrary integration constant. 
Also the dark mass can be calculated from Eq. (\ref{eqsfo17}) as,
\begin{equation}\label{clb03}
Q(r)=Q_{0}+\frac{3\alpha P_{0}}{2r}
\end{equation}
where again $Q_{0}$ is an integration constant. 
For this particular choice we have from Eqs. (\ref{eqsfo10}) and (\ref{eqsfo11}) 
$\nu '=-\lambda '$. Hence the metric elements are given by,
\begin{equation}\label{clb04}
e^{\nu}=e^{-\lambda}=1-\frac{2GM+Q_{0}}{r}-\frac{3\alpha P_{0}}{2r^{2}}
+\frac{F(\mathcal{R})-\Lambda _{4}}{3}r^{2}
\end{equation}
This solution has several interesting features which we discuss now. 
Firstly this solution is asymptotically dS (AdS) or flat 
depending on the sign of $\left(F(\mathcal{R})-\Lambda _{4} \right)$ being negative (positive) or zero. 
Then there is an analogous charge term which is the coefficient of 
$1/r^{2}$ term and is given by $-3\alpha P_{0}/2$. Finally we have a mass term 
given by, $2GM+Q_{0}$. Thus we note that the charge term is coming solely from 
the dark pressure term and thus has its origin in the bulk geometry. Similar 
argument hold true for the mass term also. However the effect of $f(\mathcal{R})$ gravity 
on the bulk actually induces a dS (AdS) nature to the vacuum solutions. 

\subsection{\textbf{Case-IV.}$\mathbf{~U+2P=0}$}

Here we consider a different condition on the dark radiation and 
dark pressure terms. In this case Eq. (\ref{eqsfo16}) leads to the 
expression for the dark mass $Q$ as,
\begin{equation}\label{clb05}
Q=\frac{2r}{3}-2GM
\end{equation}
along with the the solution for dark radiation 
term and dark pressure term as,
\begin{equation}\label{clb08}
U(r)=-2P(r)=\frac{2}{9\alpha r^{2}}
\end{equation}
The metric elements in this case can be evaluated as,
\begin{eqnarray}
e^{-\lambda}&=&\frac{1}{3}+\frac{F(\mathcal{R})-\Lambda _{4}}{3}r^{2}
\label{clb06}
\\
e^{\nu}&=&C_{0}r^{2}
\label{clb07}
\end{eqnarray}
Note that this solution actually represents a naked singularity since 
the event horizon is determined by the equation, $e^{\nu}=0$. Thus though
the $f(\mathcal{R})$ model modifies the $e^{\lambda}$ term 
however it yields a naked singularity solution. Moreover $e^{-\lambda}=0$ determines the 
null surface, however in this situation the null surface exists only if 
$\Lambda _{4}>F(\mathcal{R})$ and is located at, $r_{h}=\sqrt{\Lambda _{4}-F(\mathcal{R})}$. 
Hence by imposing 
appropriate conditions we obtain either black hole solution with event horizon or solution with  
naked singularity. 

In this context we should mention that naked singularities are just not some artifact, 
these can be used to probe structures as well. For example we can use naked singularity to 
take part in gravitational lensing and time delay, with centroid deformation of astrophysical objects 
\cite{Bhadra2002,Bhadra2008}.

\section{Stability of The Solutions}\label{Vacfstab}

Stability of black holes under linearized perturbation is considered as an important problem in black hole 
physics. Here we  consider gravitational perturbation in a 
static spherically symmetric background. Gauge invariant formalisms were 
developed in an arbitrary static background metric having the form $-g_{tt}=g^{rr}=f(r)$. 
It turns out that for certain ranges of the parameter space the Hamiltonian is positive guarantying the 
self-adjoint extension of it under suitable boundary condition. 

The perturbation can be grouped into three types: scalar, vector and tensor perturbations. Expansion of 
each of these perturbations in harmonic functions leads to a set of equations expressed in terms of gauge 
covariant variables. Further reduction of these equations then reduces them to a set of 
decoupled wave equation in the form: 
\begin{equation}
\left(\square -\frac{1}{f(r)}V\right)\Phi =0
\end{equation}
where as usual, $\square$ represents the d'Alembertian operator with respect to the two dimensional metric. 
Also $\Phi =\Phi _{S},\Phi _{V}$ and $\Phi _{T}$ represent scalar, vector and tensor perturbations 
respectively. The potential function for each of these perturbation modes 
corresponds to \cite{Kodama2003a}:
\begin{align}
V_{T}&=\frac{f(r)}{r^{2}}\left(r\dfrac{df(r)}{dr}+\ell (\ell +1)\right)
\\
V_{V}&=\frac{f(r)}{r^{2}}\left(2f(r)-r\dfrac{df(r)}{dr}+(\ell -1) (\ell +2)\right)
\\
V_{S}&=\frac{f(r)U(r)}{16r^{2}\left(m+3x\right)^{2}}
\end{align}
where we have used the following expressions:
\begin{align}
U(r)=144 x^{3}+144 mx^{2}+48mx+16m^{3}
\\
x\equiv 1-f(r),\qquad m\equiv (\ell -1)(\ell +2)
\end{align}
It should be noted that the total number of independent components of the scalar, vector and tensor modes adds up to 2, the number of independent degrees of freedom for 
graviton in the brane. 
Since the tensor mode has no degrees of freedom we need to concentrate only on the vector and scalar modes. 

Let us now consider the black hole and naked singularity solutions obtained 
in the previous section using effective gravitational field equation on the brane. Most of these solutions 
are quiet complex and we shall focus into some appropriate limiting cases.
\begin{itemize}

\item We start with the choice of vanishing dark radiation i.e. $U=0$. From the previous section it 
is evident that in general the solution is complex and not in closed form. Thus we consider the limit 
$F(\mathcal{R})\rightarrow \Lambda _{4}$, where from Eq. (\ref{clb09}) it is evident that this leads to 
Schwarzschild form for $e^{-\lambda}$. However in this limit $e^{\nu}$ becomes $(1-2M/r)$ with some 
correction factors of $\mathcal{O}(\alpha)$. Thus in the small $\alpha$ limit the solution is Schwarzschild 
in nature. Hence all the potentials $V_{T}$, $V_{V}$ and $V_{S}$ are positive implying existence of 
self adjoint operators and hence the stability. Thus for small $F(\mathcal{R})-\Lambda _{4}$ the 
deviation from Schwarzschild solution would indeed be small resulting into stability of the solution. 

\item Next we discuss the case of vanishing dark pressure. In this case the solutions are not exact 
and even the zeroth order solution for $e^{-\lambda}$ looks like Schwarzschild de-Sitter. However the 
other one is merely a constant. Thus from the expressions for the potential it turns out they depend on 
the $e^{-\lambda}$ at the outside and thus will represent stable solution for the range of parameter 
space where $f(r)>0$. From large $r$ limit we observe that stability requires the condition $A_{2}>0$, 
which is acceptable since this in turn implies that dark radiation to be positive from Eq. (\ref{new02a}). 
Thus positivity of the dark radiation term ensures stability of this solution at zeroth order. Since 
we have higher order solutions in a perturbative form, the stability of the full solution is expected to  
be dominated by the zeroth order term.

\item The most important case in our hand is the situation where dark pressure and dark radiation satisfies 
the constraint relation $2U+P=0$. In this case we can determine stability exactly. 
For this solution it turns out that $V_{T}$ and $V_{S}$ are positive for all choices of 
$F(\mathcal{R})-\Lambda _{4}$. However though $V_{V}$ is positive for $F(\mathcal{R})>\Lambda _{4}$ 
it becomes negative for the other choice. Hence All these modes are positive ensuring stability of the 
solution for the parameter space: $F(\mathcal{R})>\Lambda _{4}$. Otherwise, the solution is though stable 
under the tensor and scalar perturbations, is not so under vector perturbation. 

\item Another important aspect of this solution comes into picture when $P_{0}=0$. Then the solution 
represents a Schwarzschild (A)de-Sitter spacetime, which under proper limit leads to the Nariai 
spacetime. This has the peculiar property that a black hole in Nariai spacetime has increasing 
surface area due to quantum corrections as shown by Bousso and Hawking 
\cite{Bousso1998,Nojiri1999,Moon2011}. This phenomenon of antievaporation was then generalized for Nariai 
black holes in $f(\mathcal{R})$ gravity \cite{Nojiri2013}, 
with $f(\mathcal{R})$ gravity playing the role of anomaly induced 
effective action leading to anti evaporation. In our case as well with $P_{0}=0$, we have Nariai black 
hole as one limit and thus our solutions will also exhibit anti-evaporation. However for $P_{0}\neq 0$, 
our solution cannot be reduced to the Nariai form and thus in general the solution presented here will not 
exhibit anti-evaporation. 

\item Finally we consider the solution which corresponds to the other constraint relation with $U+2P=0$. In this case we have both 
black hole and naked singularity depending on $\Lambda _{4}>F(\mathcal{R})$ or otherwise. In this case at 
large $r$ limit both the solutions can be taken as $1+Cr^{2}$. It turns out that, $V_{T}$ and $V_{V}$ 
are positive for all choices between $F(\mathcal{R})$ and $\Lambda _{4}$, however $V_{V}$ ensures stability 
for the black hole solution not for the naked singularity. Thus the black hole solution is stable 
under all these perturbation, while the global naked singularity is stable only under tensor and scalar 
perturbation, but not under vector perturbation. 

\end{itemize}
Thus we observe that the solutions present here are mostly stable under perturbations, 
except in some specific cases where the vector mode of the perturbation shows instability. Also 
we have pointed out that our solution reduces to the Nariai form and thus exhibits anti-evaporation 
in $f(\mathcal{R})$ gravity, similar to previously obtained results.

\section{Static spherically symmetric brane with conformal motion}\label{statcon}

We can use symmetries to explore the connection between geometry 
and matter through Einstein's equation. The most important of such 
symmetries can be realized through the use of conformal 
Killing vectors. The symmetry under which the 
space-time manifold admits conformal Killing vectors are known as, 
conformal motion. In this section we derive a particular metric which admits conformal motions. 
For the spherically symmetric and static solutions 
on the brane if one requires to have 
one-parameter group of conformal motion, the following condition results,
\begin{equation}\label{scon01}
\mathcal{L}_{\xi}h_{\mu \nu}=\xi _{\mu ;\nu}+\xi _{\nu ;\mu}=\phi (r)g_{\mu \nu}
\end{equation}
In the above relation $\xi$ is the conformal Killing vector and $\phi(r)$ is 
the conformal factor, while the above symmetry of the metric is known as conformal 
motion.
The above relation should hold for all the individual metric components. 
In this relation $h_{\mu \nu}$ is the metric determining 
the vacuum space-time configuration, $\xi _{\mu}$ is a vector field 
in this space-time with respect to which the Lie variation 
has been taken and $\phi (r)$ is an arbitrary function 
of the radial coordinate. Then following the procedure adopted in 
Ref. \cite{Herrera} to determine interior structure of stellar objects, 
here also we can impose some symmetry requirement like, $\xi ^{\mu}u_{\mu}=0$.  
This symmetry enables one to determine all 
the unknowns exactly using the effective Einstein's equation. 
Thus using the metric ansatz given by Eq. (\ref{eqsfo9}), 
the above equation is shown to be equivalent to \cite{Herrera},
\begin{eqnarray}\label{scon02}
e^{\nu}=A^{2}r^{2}
\nonumber
\\
\phi (r)=Ce^{-\lambda /2}
\\
\xi ^{\mu}=D\delta _{0}^{\mu}+\frac{\phi r}{2}\delta _{1}^{\mu}
\nonumber
\end{eqnarray}
where $A$, $C$ and $D$ are arbitrary constants. 
With the above results the Einstein equations 
(\ref{eqsfo10}), (\ref{eqsfo11}) and (\ref{eqsfo12}) reduce to,
\begin{eqnarray}
\frac{1}{r^{2}}\left[1-\frac{\phi ^{2}(r)}{C^{2}} \right]
-\frac{2\phi \phi '}{rC^{2}}=3\alpha U-\left[F(\mathcal{R})-\Lambda _{4} \right]
\label{scon03}
\\
\frac{1}{r^{2}}\left(1-3\frac{\phi ^{2}}{C^{2}}\right)=
-\alpha \left(U+2P\right)-\left[F(\mathcal{R})-\Lambda _{4}\right]
\label{scon04}
\\
\frac{1}{C^{2}}\frac{\phi ^{2}}{r^{2}}+\frac{2}{C^{2}}\frac{\phi \phi '}{r}=
\alpha \left(U-P\right) +\left(F(\mathcal{R})-\Lambda _{4} \right)
\label{scon05}
\end{eqnarray}
From Eqs. (\ref{scon04}) and (\ref{scon05}) we obtain the 
dark radiation and dark pressure in terms of the unknown function $\phi$ as,
\begin{eqnarray}
P(r)=-\frac{1}{3\alpha}\left[\frac{2}{C^{2}}\frac{\phi \phi '}{r}
+\frac{1}{r^{2}}\left(1-2\frac{\phi ^{2}}{C^{2}} \right) \right]
\label{scon06}
\\
U(r)=\frac{1}{3\alpha}\left[\frac{4}{C^{2}}\frac{\phi \phi '}{r}
-\frac{1}{r^{2}}\left(1-5\frac{\phi ^{2}}{C^{2}}\right)
-3\left(F(\mathcal{R})-\Lambda _{4} \right) \right]
\label{scon07}
\end{eqnarray}
Then from Eq. (\ref{scon03}) and the expression for dark radiation, 
the differential equation satisfied by $\phi (r)$ turns out to be,
\begin{equation}\label{scon08}
\frac{3}{C^{2}}\phi \phi '=\frac{1}{r}
\left(1-3\frac{\phi ^{2}}{C^{2}} \right)
+4r\left(F(\mathcal{R})-\Lambda _{4} \right)
\end{equation}
This can be solved with little 
effort to yield the general solution as,
\begin{equation}\label{scon09}
\phi ^{2}=\frac{C^{2}}{3}\left[1+\frac{B}{r^{2}}
+2\left(F(\mathcal{R})-\Lambda _{4} \right)r^{2} \right]
\end{equation}
where, $B$ is an integration constant. Thus full solution corresponding to 
this one parameter symmetry group of conformal motion leads to,
\begin{eqnarray}
e^{\nu}=A^{2}r^{2}
\label{scon10}
\\
e^{-\lambda}=\frac{1}{3}\left[1+\frac{B}{r^{2}}
+2\left(F(\mathcal{R})-\Lambda _{4} \right)r^{2} \right]
\label{scon11}
\\
U(r)=\frac{1}{9\alpha r^{2}}\left[2+\frac{B}{r^{2}}
+9\left(F(\mathcal{R})-\Lambda _{4} \right)r^{2} \right]
\label{scon12}
\\
P(r)=\frac{1}{9\alpha r^{2}}\left[-1+\frac{4B}{r^{2}} \right]
\end{eqnarray}
There exists another important properties of the field equations. 
Having obtained a single solution we can make a 
transformation such that, $r\rightarrow \bar{r}(r)$, 
$U\rightarrow \bar{U}(U)$, $P\rightarrow \bar{P}(P)$ and 
$Q\rightarrow \bar{Q}(Q)$ \cite{Collins77}, called homology transformations. 
The homology properties of the equations determining dark radiation and 
dark pressure can be simplified by assuming $\gamma =P(U)/U=\textrm{constant}$ 
and $c_{s}=dP/dU=\textrm{constant}$. 
The above transformations are being generated with the infinitesimal generator as, 
$\hat{L}=\zeta (r)\partial /\partial r+\psi ^{1}(U)\partial /\partial U
+\psi ^{2}(Q)\partial/\partial Q$. Then 
in order to have consistent solutions we must have, 
$\zeta =0$, $\psi ^{1}=U$ and $\psi ^{2}=Q+2GM$. Thus with inclusion of 
$F(\mathcal{R})$ gravity the infinitesimal generator for the 
homologous transformation becomes restricted compared to that in Einstein 
gravity. 

\section{Some Thermodynamic Features}\label{Vacftherm}

In this section we will discuss thermodynamics associated with 
these spherically symmetric vacuum spacetime. 
Our main motive is to observe if there exists any 
thermodynamic interpretation which is induced solely by the bulk. 
We focus on the line element obtained for the condition $2U+P=0$ 
which has the following expression,
\begin{eqnarray}\label{therm01}
ds^{2}&=&-\left(1-\frac{2GM+Q_{0}}{r}-\frac{3\alpha P_{0}}{2r^{2}}
+\frac{F(\mathcal{R})-\Lambda _{4}}{3}r^{2}\right)dt^{2}
\nonumber
\\
&+&
\left(1-\frac{2GM+Q_{0}}{r}-\frac{3\alpha P_{0}}{2r^{2}}+
\frac{F(\mathcal{R})-\Lambda _{4}}{3}r^{2}\right)^{-1}dr^{2}
+r^{2}d\Omega ^{2}
\end{eqnarray}
The horizon is determined by setting coefficient of $g_{tt}$ to zero, 
which in turn leads to the equation,
\begin{equation}\label{therm02}
1-\frac{2GM+Q_{0}}{r}-\frac{3\alpha P_{0}}{2r^{2}}+
\frac{F(\mathcal{R})-\Lambda _{4}}{3}r^{2}=0
\end{equation}
Then the mass term equivalent to internal energy of a thermodynamic system 
can be obtained in terms of the horizon radius as,
\begin{equation}\label{therm03}
M(r_{h})=\frac{r_{h}}{2}-\frac{Q_{0}}{2}-\frac{3\alpha P_{0}}{4r}+
\frac{F(\mathcal{R})-\Lambda _{4}}{6}r^{3}
\end{equation}
The surface area of the event horizon is given by, 
$A=\pi r_{h}^{2}$, while the entropy for the black hole 
is given by, $S=k_{B}A/4\hbar=k_{B}\pi r_{h}^{2}/4\hbar r$. 
Choosing $\hbar=1$ and Boltzmann constant appropriately 
we readily obtain,
\begin{equation}\label{therm04}
S=r_{h}^{2}
\end{equation}
Thus the mass of the black hole in terms of the entropy becomes,
\begin{equation}\label{therm05}
M(S)=\frac{\sqrt{S}}{2}-\frac{Q_{0}}{2}-\frac{3\alpha P_{0}}{4\sqrt{S}}+
\frac{F(\mathcal{R})-\Lambda _{4}}{6}S^{3/2}
\end{equation}
This leads to the first law of black hole mechanics as,
\begin{equation}\label{therm06}
dM=TdS+\Phi d(F(\mathcal{R})-\Lambda _{4})
\end{equation}
from which the black hole temperature turns out to be:
\begin{equation}\label{therm07}
T=\frac{1}{4\sqrt{S}}+\frac{3\alpha P_{0}}{8S^{3/2}}+\frac{F(\mathcal{R})-\Lambda _{4}}{4}\sqrt{S}
\end{equation}
while the chemical potential has the following 
expression:
\begin{equation}\label{therm08}
\Phi =\frac{S^{3/2}}{6}
\end{equation}
From the expression of temperature as a function of entropy it turns out that 
the specific heat has the following behavior:
\begin{eqnarray}\label{therm09}
C_{V}=T\left(\frac{\partial S}{\partial T}\right)_{F(\mathcal{R})-\Lambda _{4}}=
\frac{\frac{1}{4\sqrt{S}}+\frac{3\alpha P_{0}}{8S^{3/2}}+\frac{F(\mathcal{R})-\Lambda _{4}}{4}\sqrt{S}}
{\frac{F(\mathcal{R})-\Lambda _{4}}{8\sqrt{S}}-\frac{1}{8S^{3/2}}-\frac{9\alpha P_{0}}{16S^{5/2}}}
\end{eqnarray}

\begin{figure*}
\begin{center}

(a)
\includegraphics[height=2in,width=2.8in]{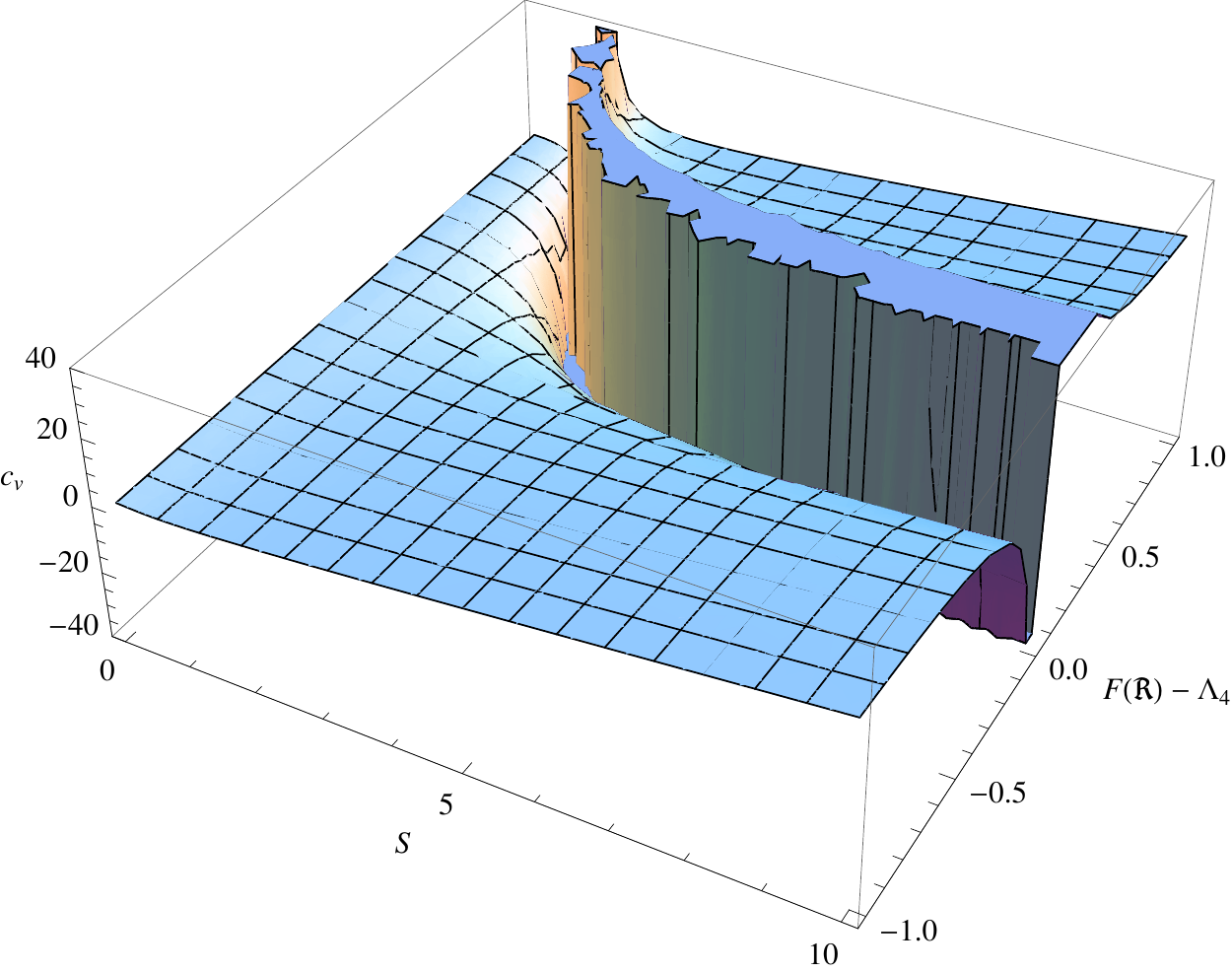}~~~
(b)
\includegraphics[height=2in,width=2.8in]{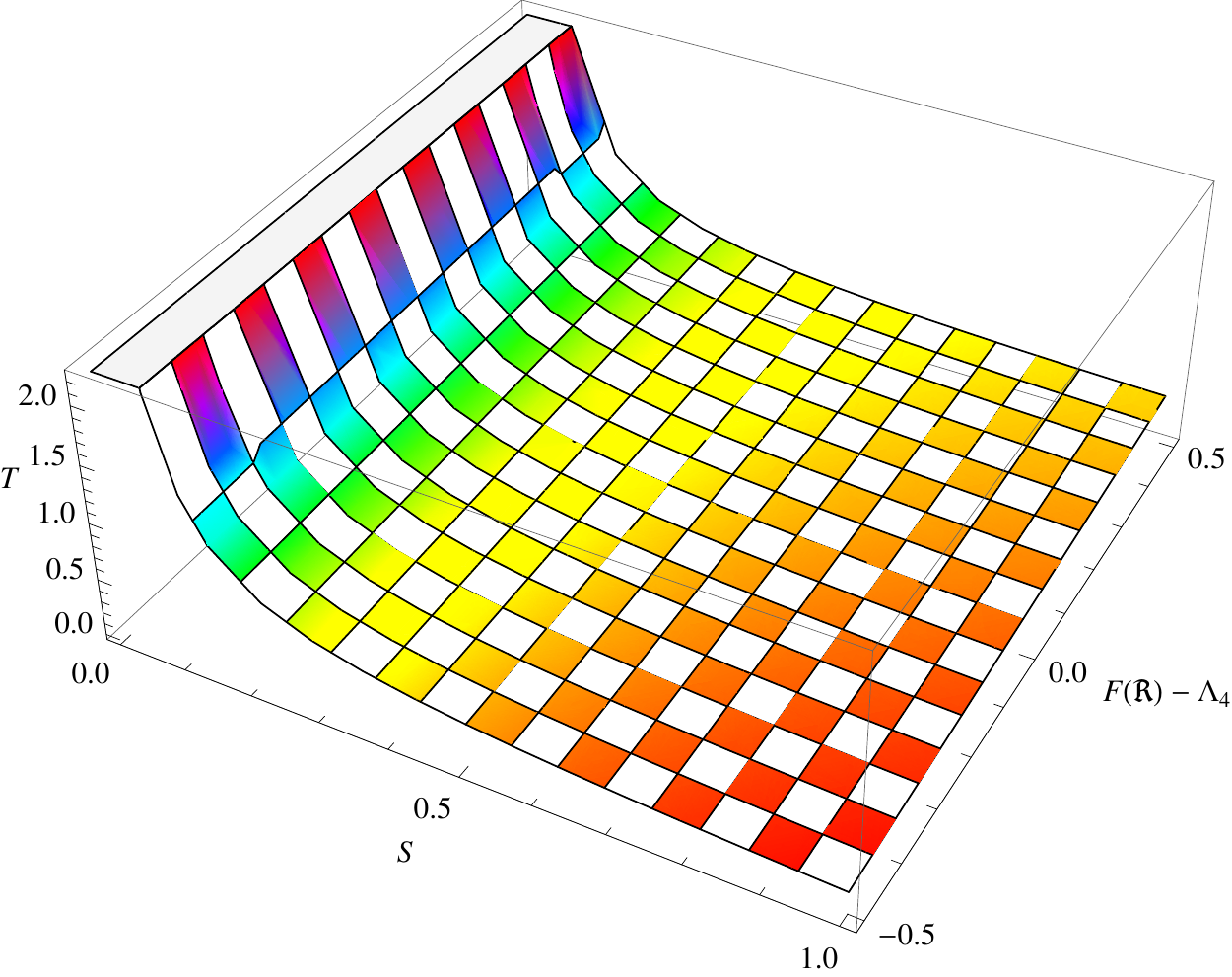}\\
(c)
\includegraphics[height=2in,width=2.8in]{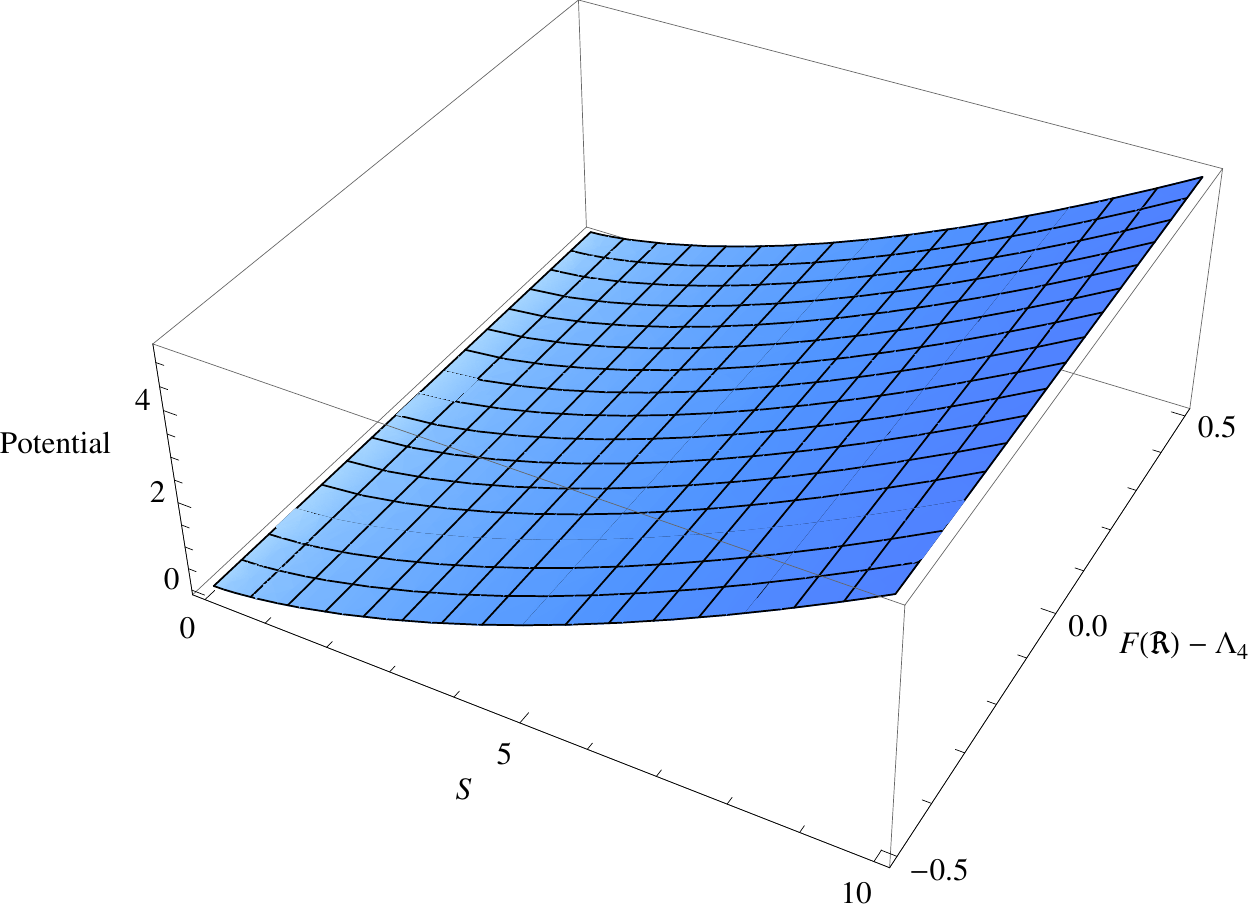}

\caption{The above figures show variation of three thermodynamic 
quantities: (a) specific heat, (b) temperature 
(c) potential with entropy and $F(\mathcal{R})-\Lambda _{4}$. 
Figure (a) clearly shows 
the existance of phase 
transition in this black hole spacetime through 
the discontinuity and divergence of the specific heat on some surface in 
entropy and $F(\mathcal{R})-\Lambda _{4}$. While continuity of both 
temperature and thermodynamic potential in (b) and (c) 
show that this phase transition is of second order.}
\label{fig1}
 
\end{center}
\end{figure*}
Figure \ref{fig1} shows that while temperature $T$ and potential $\phi$ are continuous with 
both the entropy and the quantity $F(\mathcal{R})-\Lambda _{4}$, specific heat shows discontinuity 
indicating a second order phase transition. The surface of discontinuity 
in the specific heat is given by,
\begin{equation}\label{therm10}
F(\mathcal{R})-\Lambda _{4}=\frac{1}{S}+\frac{9\alpha P_{0}}{2S}
\end{equation}
In order to understand the physics behind these results, it is always illuminating to 
discuss some limiting cases. For example, if we assume pure Einstein gravity, 
where $F(\mathcal{R})=0$, then with the assumption of $\Lambda _{4}\sim 0$, we arrive at the 
Reissner-Nordstr\"{o}m solution. From Eq. (\ref{therm09}) the specific heat turns out to be 
$C_{v}=-S\left(6\alpha P_{0}+4S\right)/\left(2S+9\alpha P_{0}\right)$. This can also be divergent, 
provided the entropy satisfies the criteria: $S=-\left(9\alpha P_{0}/2\right)$. 
In general $P_{0}$ is taken to be positive and thus the above relation cannot be satisfied in general. 
Hence the bulk term with positive dark pressure cannot lead to second order phase transition. However 
the other limit is interesting. For $P_{0}=0$, we get the divergence of specific heat to correspond to 
the condition: $S=1/\left(F(\mathcal{R})-\Lambda _{4}\right)$. Thus for $F(\mathcal{R})>\Lambda _{4}$ we 
have second order phase transition. Our calculations therefore confirm that Schwarzschild Anti-de Sitter 
solution shows second order phase transition.

The case for which 
the dark radiation vanishes, i.e. $U=0$ also exhibits the appearance of black hole horizon. The actual calculations are quiet complex, and we have 
presented them in App. \ref{AppA}. However here we consider some limiting cases and discuss the corresponding thermodynamic features. 
The first case corresponds to $F(\mathcal{R})-\Lambda _{4}=0$. In this situation the 
solution for the metric elements resemble Schwarzschild solution with 
no associated phase transition. We cannot take $P=0$ as in that case the metric elements would diverge. 
Thus another obvious choice is $M=0$. Then also horizon appears and the specific heat diverges  
for $\Lambda _{4}>F(\mathcal{R})$. Thus this configuration exhibits an 
opposite effect in respect to $2U+P=0$ case.

We therefore observe that in both the black hole solutions the specific heat diverge showing 
second order phase transition, due to the presence of $F(\mathcal{R})$ gravity in the bulk. 
Thus bulk $F(\mathcal{R})$ gravity plays a crucial 
role in determining the thermodynamic feature of the brane world black holes.

\section{Discussion}\label{Vacfdis}

In this work we have considered a bulk action with a $f(\mathcal{R})$ term, 
where $\mathcal{R}$ is the bulk curvature. 
Starting from the bulk action we have derived the full effective 
Einstein's equation on the brane located at $y=0$, which under 
$f(\mathcal{R})\rightarrow \mathcal{R}$ limit goes to the usual 
Gauss Codazzi equation in Einstein gravity. In order to get 
spherically symmetric solutions we have assumed that in the 
region of interest there is no matter field present on the brane 
and also the four dimensional scalar curvature is 
constant. Under these conditions the Einstein equation simplifies 
considerably, however the Weyl tensor on bulk has 
non trivial decomposition on the brane leading to the appearance of dark pressure and 
dark radiation in the effective Einstein's equation. Also 
the induced four dimensional cosmological constant and contribution 
from $f(\mathcal{R})$ term have significant effects on 
the solutions of the effective Einstein's equation on the brane.

Due to the presence of $f(\mathcal{R})$ gravity in the bulk, 
Einstein's equation on the brane picks up an extra contribution 
which acts as an effective cosmological constant having expression: 
$F(\mathcal{R})-\Lambda _{4}$. Thus though the four dimensional 
parameter $\Lambda _{4}$ is not small, an effective small 
cosmological constant can be generated by fine tuning 
$\Lambda _{4}$ and $F(\mathcal{R})$. Hence we can argue that 
the observed smallness of four dimensional cosmological 
constant is due to a fine tuning of induced cosmological 
constant on the brane with the $f(\mathcal{R})$ term in the bulk. 

From the effective Einstein's equation we can solve for the 
metric elements as well as for dark radiation and dark pressure term 
provided a relation between dark pressure and dark radiation 
term is assumed. For four such choices 
the equations get sufficiently simplified such that analytic solutions can be obtained. 
We have derived all the metric elements for these four choices. Among the four solutions 
two of them show the presence of event horizon and thus is important 
from thermodynamic point of view. 
On the other hand the other two solutions lead to naked singularity and thus 
does not have much astrophysical importance. The important 
features of these solutions are the asymptotic non-flatness 
due to presence of $f(\mathcal{R})$ term. This might be of some relevance in the context of 
AdS-CFT correspondence. 

After obtaining various solutions leading to either a black hole or a naked singularity, we 
have performed a stability analysis of our solutions in some appropriate limit. It turns out that 
the solutions are stable under tensor and scalar perturbations, while under certain choices of parameters 
the vector mode leads to instability. Also some solutions can be reduced to Nariai form, where 
the well known anti-evaporation in $f(\mathcal{R})$ gravity takes place leading to an increase in 
the area of the event horizon. However we have argued that in general the  
solutions are stable under perturbations. 

In order to get some idea about solutions representing stellar interior, 
a symmetry transformation, known as conformal motion 
is invoked. For this particular symmetry class we can solve the 
field equations exactly. This leads to direct evaluation of dark 
pressure and radiation using these symmetries. Also there exists 
another class of transformations known as homology transformations. 
For this class of solutions the homology operator has been evaluated 
and it turns out that $f(\mathcal{R})$ term makes 
the homology class restricted compared to that in Einstein gravity. 

Finally we consider thermodynamical behavior of these spherically symmetric space-times. 
Since thermodynamics is intimately 
connected to existence of a horizon, we consider only the two relevant cases. 
Here also the $f(\mathcal{R})$ term plays a dominant role in determining 
the thermodynamic behavior. In both the cases, 
the temperature and chemical potentials are found to be continuous, 
while the specific heat turns out to be  discontinuous 
along a surface indicating a second order phase transition. 
Such features of these spherically symmetric solutions have their origin 
in the $f(\mathcal{R})$ term in the bulk action  
and only because of the presence of higher curvature terms in the action, the black hole solutions 
exhibit a phase transition, which, is second order in nature. 

\section*{Acknowledgements}

S.C. is funded by a SPM fellowship from CSIR, Government of India. 
\appendix

\section{Some Detailed Expressions}\label{AppA}
Here we present general expressions for various thermodynamic 
quantities for the case $U = 0$ which have been discussed in Sec.III.

Under the condition of vanishing dark radiation also we have a horizon structure to our solution. Therefore 
we can work out the thermodynamic features.  In this case, the horizon radius turns out 
to be in terms of the mass $M$ and the parameter $F(\mathcal{R})-\Lambda _{4}$ 
with unit $G=1$ as:
\begin{eqnarray}\label{therm11}
r_{1}=\frac{3^{-2/3}\left(F(\mathcal{R})-\Lambda _{4}\right)
+\left(-M\left(F(\mathcal{R})-\Lambda _{4}\right)^{2}
+\sqrt{3}\sqrt{\frac{\left(F(\mathcal{R})-\Lambda _{4}\right)^{3}}{27}
+\left[-1+9\left(F(\mathcal{R})-\Lambda _{4}\right)M^{2}\right]}\right)^{2/3}}
{3^{-1/3}\left(F(\mathcal{R})-\Lambda _{4}\right)\left(-M\left(F(\mathcal{R})-\Lambda _{4}\right)^{2}
+\sqrt{3}\sqrt{\frac{\left(F(\mathcal{R})-\Lambda _{4}\right)^{3}}{27}
+\left[-1+9\left(F(\mathcal{R})-\Lambda _{4}\right)M^{2}\right]}\right)^{1/3}}
\end{eqnarray}
Then by the previous conditions: $\hbar =1$ and an appropriate choice of Boltzmann constant we get entropy 
to be $S=r_{h}^{2}$. From the first law of black hole mechanics as 
presented in Eq. (\ref{therm06}) the temperature 
turns out to be,
\begin{eqnarray}\label{therm12}
T^{-1}&=&\left(\frac{\partial S}{\partial M}\right)_{F(\mathcal{R})-\Lambda _{4}}
\nonumber
\\
&=&\frac{2r_{h}}{3}\frac{\left(-M\left(F(\mathcal{R})-\Lambda _{4}\right)^{2}
+\sqrt{3}\sqrt{\frac{\left(F(\mathcal{R})-\Lambda _{4}\right)^{3}}{27}
+\left[-1+9\left(F(\mathcal{R})
-\Lambda _{4}\right)M^{2}\right]}\right)^{2/3}-3^{-2/3}
\left(F(\mathcal{R})-\Lambda _{4}\right)}
{\left(-M\left(F(\mathcal{R})-\Lambda _{4}\right)^{2}
+\sqrt{3}\sqrt{\frac{\left(F(\mathcal{R})-\Lambda _{4}\right)^{3}}{27}
+\left[-1+9\left(F(\mathcal{R})-\Lambda _{4}\right)M^{2}\right]}\right)^{4/3}}
\nonumber
\\
&\times&\frac{1}{3^{-1/3}\left(F(\mathcal{R})-\Lambda _{4}\right)}
\left[-\left(F(\mathcal{R})-\Lambda _{4}\right)^{2}
+\frac{9\sqrt{3}M\left(F(\mathcal{R})-\Lambda _{4}\right)}
{\sqrt{\left(F(\mathcal{R})-\Lambda _{4}\right)^{3}/27
+\left(-1+9M^{2}\left(F(\mathcal{R})-\Lambda _{4}\right)\right)}}\right]
\end{eqnarray}
while the potential $\phi$ can be obtained by solving the equation:
\begin{eqnarray}\label{therm14}
0&=&\left[3^{-1/3}\left(F(\mathcal{R})-\Lambda _{4}\right)
\left(-M\left(F(\mathcal{R})-\Lambda _{4}\right)^{2}
+\sqrt{3}\sqrt{\frac{\left(F(\mathcal{R})-\Lambda _{4}\right)^{3}}{27}
+\left[-1+9\left(F(\mathcal{R})-\Lambda _{4}\right)M^{2}\right]}\right)^{1/3}\right]^{-1}
\nonumber
\\
&\times&\Big[ 3^{1/3}+\frac{2}{3}\left(-M\left(F(\mathcal{R})-\Lambda _{4}\right)^{2}
+\sqrt{3}\sqrt{\frac{\left(F(\mathcal{R})-\Lambda _{4}\right)^{3}}{27}
+\left[-1+9\left(F(\mathcal{R})-\Lambda _{4}\right)M^{2}\right]}\right)^{-1/3} 
\nonumber
\\
&\times&\left\lbrace-3\phi \left(F(\mathcal{R})-\Lambda _{4}\right)^{2}
-6M\left(F(\mathcal{R})-\Lambda _{4}\right) +\frac{\sqrt{3}}{2}
\frac{\left(F(\mathcal{R})-\Lambda _{4}\right)^{2}/3+27M^{2}
+54M\left(F(\mathcal{R})-\Lambda _{4}\right)\phi}
{\sqrt{\frac{\left(F(\mathcal{R})-\Lambda _{4}\right)^{3}}{27}
+\left[-1+9\left(F(\mathcal{R})-\Lambda _{4}\right)M^{2}\right]}} \right\rbrace \Big]
\nonumber
\\
&-&\left[3^{-2/3}\left(F(\mathcal{R})-\Lambda _{4}\right)+\left(-M\left(F(\mathcal{R})-\Lambda _{4}\right)^{2}
+\sqrt{3}\sqrt{\frac{\left(F(\mathcal{R})-\Lambda _{4}\right)^{3}}{27}
+\left[-1+9\left(F(\mathcal{R})-\Lambda _{4}\right)M^{2}\right]}\right)^{2/3}\right]
\nonumber
\\
&\times&\left[3^{-1/3}\left(F(\mathcal{R})-\Lambda _{4}\right)
\left(-M\left(F(\mathcal{R})-\Lambda _{4}\right)^{2}
+\sqrt{3}\sqrt{\frac{\left(F(\mathcal{R})-\Lambda _{4}\right)^{3}}{27}
+\left[-1+9\left(F(\mathcal{R})-\Lambda _{4}\right)M^{2}\right]}\right)^{1/3}\right]^{2}
\nonumber
\\
&\times&\Big[3^{2/3}\left(-M\left(F(\mathcal{R})-\Lambda _{4}\right)^{2}
+\sqrt{3}\sqrt{\frac{\left(F(\mathcal{R})-\Lambda _{4}\right)^{3}}{27}
+\left[-1+9\left(F(\mathcal{R})-\Lambda _{4}\right)M^{2}\right]}\right)
\nonumber
\\
&+&3^{-1/3}\left(F(\mathcal{R})-\Lambda _{4}\right)\left(-M\left(F(\mathcal{R})-\Lambda _{4}\right)^{2}
+\sqrt{3}\sqrt{\frac{\left(F(\mathcal{R})-\Lambda _{4}\right)^{3}}{27}
+\left[-1+9\left(F(\mathcal{R})-\Lambda _{4}\right)M^{2}\right]}\right)^{-2/3}
\nonumber
\\
&\times&\left\lbrace-3\phi \left(F(\mathcal{R})-\Lambda _{4}\right)^{2}
-6M\left(F(\mathcal{R})-\Lambda _{4}\right) +\frac{\sqrt{3}}{2}
\frac{\left(F(\mathcal{R})-\Lambda _{4}\right)^{2}/3+27M^{2}
+54M\left(F(\mathcal{R})-\Lambda _{4}\right)\phi}
{\sqrt{\frac{\left(F(\mathcal{R})-\Lambda _{4}\right)^{3}}{27}
+\left[-1+9\left(F(\mathcal{R})-\Lambda _{4}\right)M^{2}\right]}} \right\rbrace \Big]
\nonumber
\end{eqnarray}
In this case the specific heat becomes,
\begin{eqnarray}\label{therm13}
C_{v}&=&T\left(\frac{\partial S}{\partial T}\right)_{F(\mathcal{R})-\Lambda _{4}}
=\left(\frac{\partial M}{\partial T}\right)_{F(\mathcal{R})-\Lambda _{4}}
\nonumber
\\
&=&\frac{2r_{h}}{3^{2/3}\left(F(\mathcal{R})-\Lambda _{4}\right)}
\Big[-\frac{1}{3^{2/3}\left(F(\mathcal{R})-\Lambda _{4}\right)}
\nonumber
\\
&+&\frac{4}{3}
\frac{\left(-M\left(F(\mathcal{R})-\Lambda _{4}\right)^{2}
+\sqrt{3}\sqrt{\frac{\left(F(\mathcal{R})-\Lambda _{4}\right)^{3}}{27}
+\left[-1+9\left(F(\mathcal{R})-\Lambda _{4}\right)M^{2}\right]}\right)^{1/3}}
{\left(-M\left(F(\mathcal{R})-\Lambda _{4}\right)^{2}
+\sqrt{3}\sqrt{\frac{\left(F(\mathcal{R})-\Lambda _{4}\right)^{3}}{27}
+\left[-1+9\left(F(\mathcal{R})
-\Lambda _{4}\right)M^{2}\right]}\right)^{2/3}-3^{-2/3}
\left(F(\mathcal{R})-\Lambda _{4}\right)}
\nonumber
\\
&-&\frac{2}{3}\frac{\left(-M\left(F(\mathcal{R})-\Lambda _{4}\right)^{2}
+\sqrt{3}\sqrt{\frac{\left(F(\mathcal{R})-\Lambda _{4}\right)^{3}}{27}
+\left[-1+9\left(F(\mathcal{R})-\Lambda _{4}\right)M^{2}\right]}\right)}
{\left\lbrace\left(-M\left(F(\mathcal{R})-\Lambda _{4}\right)^{2}
+\sqrt{3}\sqrt{\frac{\left(F(\mathcal{R})-\Lambda _{4}\right)^{3}}{27}
+\left[-1+9\left(F(\mathcal{R})-\Lambda _{4}\right)M^{2}\right]}\right)^{2/3}
-3^{-2/3}\left(F(\mathcal{R})-\Lambda _{4}\right)\right\rbrace ^{2}}
\nonumber
\\
&-&\frac{\left(-M\left(F(\mathcal{R})-\Lambda _{4}\right)^{2}
+\sqrt{3}\sqrt{\frac{\left(F(\mathcal{R})-\Lambda _{4}\right)^{3}}{27}
+\left[-1+9\left(F(\mathcal{R})-\Lambda _{4}\right)M^{2}\right]}\right)^{4/3}}
{\left(-M\left(F(\mathcal{R})-\Lambda _{4}\right)^{2}
+\sqrt{3}\sqrt{\frac{\left(F(\mathcal{R})-\Lambda _{4}\right)^{3}}{27}
+\left[-1+9\left(F(\mathcal{R})
-\Lambda _{4}\right)M^{2}\right]}\right)^{2/3}
-3^{-2/3}\left(F(\mathcal{R})-\Lambda _{4}\right)}
\nonumber
\\
&\times&\left\lbrace \frac{9\sqrt{3}\left(F(\mathcal{R})
-\Lambda _{4}\right)}{\sqrt{\left(F(\mathcal{R})-\Lambda _{4}\right)^{3}/27
+\left(-1+9M^{2}\left(F(\mathcal{R})
-\Lambda _{4}\right)\right)}}-\frac{\sqrt{3}81M^{2}
\left(F(\mathcal{R})-\Lambda _{4}\right)^{2}}
{\left(F(\mathcal{R})-\Lambda _{4}\right)^{3}/27
+\left(-1+9M^{2}\left(F(\mathcal{R})-\Lambda _{4}\right)\right)^{3/2}} \right\rbrace 
\nonumber
\\
&\times&\left[-\left(F(\mathcal{R})-\Lambda _{4}\right)^{2}
+\frac{9\sqrt{3}M\left(F(\mathcal{R})-\Lambda _{4}\right)}
{\sqrt{\left(F(\mathcal{R})-\Lambda _{4}\right)^{3}/27
+\left(-1+9M^{2}\left(F(\mathcal{R})-\Lambda _{4}\right)\right)}}\right]^{-2}\Big]^{-1}
\end{eqnarray}
It is evident from the expression of the specific heat that it diverges at the surface 
given by:
\begin{eqnarray}\label{therm14}
\frac{4}{3}&\times&\left(-M\left(F(\mathcal{R})-\Lambda _{4}\right)^{2}
+\sqrt{3}\sqrt{\frac{\left(F(\mathcal{R})-\Lambda _{4}\right)^{3}}{27}
+\left[-1+9\left(F(\mathcal{R})-\Lambda _{4}\right)M^{2}\right]}\right)^{1/3}
\nonumber
\\
&=&\left(-M\left(F(\mathcal{R})-\Lambda _{4}\right)^{2}
+\sqrt{3}\sqrt{\frac{\left(F(\mathcal{R})-\Lambda _{4}\right)^{3}}{27}
+\left[-1+9\left(F(\mathcal{R})-\Lambda _{4}\right)M^{2}\right]}\right)^{4/3}
\nonumber
\\
&\times&\left\lbrace \frac{9\sqrt{3}
\left(F(\mathcal{R})-\Lambda _{4}\right)}{\sqrt{\left(F(\mathcal{R})-\Lambda _{4}\right)^{3}/27
+\left(-1+9M^{2}\left(F(\mathcal{R})
-\Lambda _{4}\right)\right)}}-\frac{81\sqrt{3}M^{2}\left(F(\mathcal{R})
-\Lambda _{4}\right)^{2}}
{\left(F(\mathcal{R})-\Lambda _{4}\right)^{3}/27
+\left(-1+9M^{2}\left(F(\mathcal{R})-\Lambda _{4}\right)\right)^{3/2}} \right\rbrace 
\nonumber
\\
&\times&\left[-\left(F(\mathcal{R})-\Lambda _{4}\right)^{2}
+\frac{9\sqrt{3}M\left(F(\mathcal{R})-\Lambda _{4}\right)}
{\sqrt{\left(F(\mathcal{R})-\Lambda _{4}\right)^{3}/27
+\left(-1+9M^{2}\left(F(\mathcal{R})-\Lambda _{4}\right)\right)}}\right]^{-2}
\nonumber
\\
&+&\frac{\left(-M\left(F(\mathcal{R})-\Lambda _{4}\right)^{2}
+\sqrt{3}\sqrt{\frac{\left(F(\mathcal{R})
-\Lambda _{4}\right)^{3}}{27}
+\left[-1+9\left(F(\mathcal{R})
-\Lambda _{4}\right)M^{2}\right]}\right)^{2/3}
-3^{-2/3}\left(F(\mathcal{R})-\Lambda _{4}\right)}
{3^{2/3}\left(F(\mathcal{R})-\Lambda _{4}\right)}
\nonumber
\\
&+&\frac{2}{3}\frac{\left(F(\mathcal{R})-\Lambda _{4}\right)
\left(-M\left(F(\mathcal{R})-\Lambda _{4}\right)^{2}
+\sqrt{3}\sqrt{\frac{\left(F(\mathcal{R})-\Lambda _{4}\right)^{3}}{27}
+\left[-1+9\left(F(\mathcal{R})-\Lambda _{4}\right)M^{2}\right]}\right)}
{\left(-M\left(F(\mathcal{R})-\Lambda _{4}\right)^{2}
+\sqrt{3}\sqrt{\frac{\left(F(\mathcal{R})-\Lambda _{4}\right)^{3}}{27}
+\left[-1+9\left(F(\mathcal{R})-\Lambda _{4}\right)
M^{2}\right]}\right)^{2/3}-3^{-2/3}\left(F(\mathcal{R})-\Lambda _{4}\right)}
\end{eqnarray}
This again shows that the black hole solution presented by the condition of vanishing dark radiation 
has a divergent behavior on the above surface which in turn 
indicates that the black hole undergoes a second order 
phase transition on this surface.




\begin{thebibliography}{100}

\bibitem{Randall99} L. Randall and R. Sundrum, \textit{Phys. Rev. Lett.} \textbf{83}, 3370 (1999) 
hep-ph/9905221;\\
L. Randall and R. Sundrum, \textit{Phys. Rev. Lett.} \textbf{83}, 4690 (1999) 
hep-th/9906064.

\bibitem{Witten96} P. Horava and E. Witten, \textit{Nucl. Phys.} \textbf{B460}, 506 (1996).

\bibitem{Csaki2000} C. Cs\'{a}ki, M. Graesser, L. Randall and J. Terning, 
\textit{Phys. Rev. D} \textbf{62}, 045015 (2000).

\bibitem{Chakraborty2014b} S. Chakraborty and S. SenGupta, \textit{Eur. Phys. J. C} 
\textbf{74}, 3045 (2014) arXiv:1306.0805.

\bibitem{Maartens01} R. Maartens, [arXiv:0101059]

\bibitem{Maeda00} T. Shromizu, K. Maeda and M. Sasaki, \textit{Phys. Rev. D} \textbf{62}, 024012 (2000).

\bibitem{Harko01} R. Maartens, \textit{Phys. Rev. D}, \textbf{62}, 084023 (2000);\\ 
C.M. Chen, T. Harko and M.K. Mak, \textit{ibid} \textbf{64}, 044013 (2001);\\
P. Binetruy, C. Deffayet and D. Langlois, \textit{Nucl. Phys.} 
\textbf{B565}, 269 (2000) [hep-th/9905210];\\ 
M.K. Mak and T. Harko, \textit{Class. Quant. Grav.} \textbf{20}, 407 (2003);\\ 
A. Coley, \textit{Class. Quant. Grav.} \textbf{19}, L45 (2002).

\bibitem{Ghosh01} N. Dadhich and S.G. Ghosh, \textit{Phys. Lett. B} \textbf{518}, 1 (2001);\\ 
M.G. Santos, F. Vernizzi and P.G. Ferreira, \textit{Phys. Rev. D} \textbf{64}, 063506 (2001);\\ 
T. Wiseman, \textit{Class. Quantum. Grav.} \textbf{19}, 3083 (2002).

\bibitem{Djouadi2008} A. Djouadi, \textit{Phys. Rep.} \textbf{457}, 1 (2008).

\bibitem{Hundi2013} R.S. Hundi and S. SenGupta, \textit{J. Phys. G} 
\textbf{40}, 075002 (2013).

\bibitem{Chakraborty2014c} S. Chakraborty and S. SenGupta, \textit{Phys. Rev. D} 
\textbf{89}, 126001 (2014) arXiv:1401.3279.

\bibitem{Dudas2005} E. Dudas and M. Quiros, \textit{Nucl. Phys. B} 
\textbf{721}, 309 (2005). 

\bibitem{Lukas1999} A. Lukas, B.A. Ovrut, D. Waldram, \textit{Phys. Rev. D} \textbf{60}, 
086001 (1999).

\bibitem{Dadhich00} N. Dadhich, R. Maartens, P. Papadopoulos and V. Rezania, 
\textit{Phys. Lett. B} \textbf{487}, 1 (2000).

\bibitem{Germani01} C. Germani and R. Maartens, \textit{Phys. Rev. D} \textbf{64} 124010 (2001).

\bibitem{Casadio14} R. Casadio, J. Ovalle, \textit{Gen. Relt. Grav.} \textbf{46}, 1669 (2014).

\bibitem{Ovalle13} J. Ovalle and F. Linares, \textit{Phys. Rev. D} \textbf{88}, 104026 (2013).

\bibitem{Casadio12} R. Casadio and J. Ovalle, \textit{Phys. Lett. B} \textbf{715}, 251 (2012). 

\bibitem{Dadhich03} S. Shankaranarayanan and N. Dadhich, gr-qc/0306111.

\bibitem{Visser03} M. Visser and D.L. Wiltshire, \textit{Phys. Rev. D} \textbf{67}, 104004 (2003).

\bibitem{Harko04} T. Harko and M.K. Mak, \textit{Phys. Rev. D} \textbf{69}, 064020 (2004).

\bibitem{Nojiri11} S. Nojiri and S. D. Odintsov, \textit{Phys. Rept.} \textbf{505}, 59 (2011);\\ 
T. P. Sotiriou and V. Faraoni, \textit{Rev. Mod. Phys.} \textbf{82}, 451 (2010);\\ 
A. De Felice and S. Tsujikawa, \textit{Living Rev. Relativity} \textbf{13}, 3 (2010).

\bibitem{Nojiri07} S. Nojiri and S.D. Odintsov, \textit{Phys. Lett. B} \textbf{652}, 343 (2007);\\
S. Nojiri and S.D. Odintsov \textit{Phys. Lett. B} \textbf{657}, 238 (2007);\\ 
S. Nojiri and S.D. Odintsov, \textit{Phys. Rev. D} \textbf{77}, 026007 (2008);\\ 
A.D. Dolgov and M. Kawasaki, \textit{Phys. Lett. B} \textbf{573}, 1 (2003).

\bibitem{Nojiri06} S. Nojiri and S.D. Odintsov, \textit{Phys. Rev. D} \textbf{74}, 086005 (2006);\\ 
S. Nojiri and S.D. Odintsov \textit{Phys. Rev. D} \textbf{78}, 046006 (2008).

\bibitem{Bamba08} K. Bamba and S.D. Odintsov, \textit{JCAP} \textbf{0804}, 024 (2008);\\ 
A. Dobado and A. de la Cruz-Dombriz, \textit{Phys. Rev. D} \textbf{74}, 087501 (2006).

\bibitem{Star80} A.A. Starobinsky, \textit{Phys. Lett. B} \textbf{91}, 99 (1980).

\bibitem{Troisi06} S. Capozziello, V.F. Cardone and A. Troisi, \textit{JCAP}, \textbf{0608}, 001 (2006);\\ 
S. Capozziello, V.F. Cardone and A. Troisi, \textit{Mon. Not. Roy. Astron. Soc.} \textbf{375}, 1423 (2007). 

\bibitem{Corda08} C. Corda, \textit{Int. J. Mod. Phys. D}, \textbf{18}, 2275 (2009);\\ 
C. Corda, \textit{Int. J. Mod. Phys. D} \textbf{19}, 2095 (2010);\\ 
C. Corda, \textit{Astropart. Phys.} \textbf{30}, 209 (2008).

\bibitem{Silva13} T.R.P. Carames, M.E.X. Guimaraes and J.M. Hoff da Silva, \textit{Phys. Rev. D} 
\textbf{87}, 106011 (2013).

\bibitem{Bazeia13} D. Bazeia, R. Menezes, A. Yu. Petrov and A. J. da Silva, \textit{Phys. Lett. B} 
\textbf{726}, 523 (2013).

\bibitem{Haghani2012} Z. Haghani, H.R. Sepangi and S. Shahidi arXiv:1201.6448.

\bibitem{Borzou2009} A. Borzou, H.R. Sepangi, S. Shahidi and R. Yousefi, \textit{Eur. Phys. Lett.} \textbf{88}, 29001 (2009).

\bibitem{Parry05} M. Parry, S. Pichler and D. Deeg, \textit{JCAP} \textbf{04}, 014 (2005).

\bibitem{Bal10} A. Balcerzak and M.P. Dabrowski, \textit{Phys. Rev. D} \textbf{81}, 123527 (2010).

\bibitem{Silva11} J.M. Hoff da Silva and M. Dias, \textit{Phys. Rev. D} \textbf{84}, 066011 (2011).

\bibitem{Chakraborty14} S. Chakraborty and S. SenGupta, \textit{Phys. Rev. D} 
\textbf{90}, 047901 (2014) arXiv:1403.3164.

\bibitem{Regge57} T. Regge and J.A. Wheeler, \textit{Phys. Rev.} \textbf{108}, 1063 (1957).

\bibitem{Zerilli70} F. Zerilli, \textit{Phys. Rev. Lett.} \textbf{24}, 737 (1970).

\bibitem{Vish70} C.V. Vishveshwara, \textit{Phys. Rev. D} \textit{1}, 2870 (1970).

\bibitem{Kodama2003a} H. Kodama and A. Ishibashi, \textit{Prog. Theo. Phys.} \textbf{110}, 701 (2003).

\bibitem{Kodama2003b} A. Ishibashi and H. Kodama, \textit{Prog. Theo. Phys.} \textbf{110}, 901 (2003).

\bibitem{Kodama2004} H. Kodama and A. Ishibashi, \textit{Prog. Theo. Phys.} \textbf{111}, 29 (2004).

\bibitem{Giveon2004} A. Giveon, B. Kol, A. Ori and A. Sever, \textit{JHEP}, \textbf{0408}, 014 (2004).

\bibitem{Bekenstein73} J.D. Bekenstein, \textit{Phys. Rev. D} \textbf{7}, 2333 (1973).

\bibitem{Hawking73} J.M. Bardeen, B. Carter and S.W. Hawking, 
\textit{Commun. Math. Phys.} \textbf{31}, 161 (1973).

\bibitem{Hawking75} S.W. Hawking, \textit{Commun. Math. Phys.} \textbf{43}, 199 (1975).

\bibitem{Padmanabhan2005a} T. Padmanabhan, \textit{Phys. Rept.} \textbf{406}, 
49 (2005) [arXiv:0311036].

\bibitem{Wald2001} R.M. Wald, \textit{Liv. Rev. Relt.} \textbf{4}, 6 (2001) [arXiv:9912119].

\bibitem{Padmanabhan2010a} T. Padmanabhan, \textit{Rept. Prog. Phys.} \textbf{73}, 
046901 (2010) [arXiv:0911.5004].

\bibitem{Davies77} P.C.W. Davies, \textit{Rep. Prog. Phys.}, \textbf{41}, 1313 (1977).

\bibitem{Davies89} P.C.W. Davies, \textit{Class. Quant. Grav.} \textbf{6}, 1909 (1989).

\bibitem{Page83} S.W. Hawking and D.N. Page, \textit{Commun. Math. Phys.} \textbf{87}, 577 (1983).

\bibitem{Hut77} P. Hut, \textit{Mon. Not. Roy. Astron. Soc.} \textbf{180}, 379 (1977).

\bibitem{York86} J.W. York, \textit{Phys. Rev. D} \textbf{33}, 2092 (1986).

\bibitem{Myung07} Y.S. Mayung, \textit{Phys. Lett. B} \textbf{645}, 639 (2007).

\bibitem{Chakraborty10} R. Biswas and S. Chakraborty, 
\textit{Gen. Relt. Gravit.} \textbf{42}, 1311 (2010);\\ 
R. Biswas and S. Chakraborty, \textit{Astrophys. Space. Sci.} \textbf{332}, 171 (2011).

\bibitem{Sengupta13} S. Choudhury and S. Sengupta, arXiv:1306.0492.

\bibitem{Bhadra2002} K.S. Virbhadra and G.F.R. Ellis, \textit{Phys. Rev. D} \textbf{65}, 103004 (2002).

\bibitem{Bhadra2008} K.S. Virbhadra and C.R. Kenton, \textit{Phys. Rev. D} \textbf{77}, 124014 (2008).

\bibitem{Bousso1998} R. Bousso and S.W. Hawking, \textit{Phys. Rev. D} \textbf{57}, 2436 (1998).

\bibitem{Nojiri1999} S. Nojiri and S.D. Odinstov, \textit{Int. J. Mod. Phys. A} \textbf{14}, 1293 (1999).

\bibitem{Moon2011} T. Moon, Y.S. Myung and E.J.Son, \textit{Eur. Phys. J. C} \textbf{71} 1777 (2011).

\bibitem{Nojiri2013} S. Nojiri and S.D. Odinstov, arXiv:1301.2775.

\bibitem{Herrera} L. Herrera and J. Ponce de Leon, \textit{J. Math. Phys.} \textbf{26}, 2303 (1985).

\bibitem{Collins77} C.B. Collins, \textit{J. Math. Phys.} \textbf{18}, 1374 (1977).

\end{thebibliography}
\end{document}